\documentstyle[twocolumn,prl,aps,epsf,floats]{revtex}

\newcommand{\simge}{\hspace*{0.2em}\raisebox{0.5ex}{$>$}
     \hspace{-0.8em}\raisebox{-0.3em}{$\sim$}\hspace*{0.2em}}

\begin{document}

\title
{
  \hfill{\normalsize DOE/ER/40561-28-INT98}\\ 
  \hfill{\normalsize TRI-PP-98-17}\\
  \hfill{\normalsize KRL MAP-235}\\
  \hfill{\normalsize NT@UW-98-24}\\[1cm]
 Renormalization of the Three-Body System
 with Short-Range Interactions}

\author
{ P.F. Bedaque$^{a,}$\thanks{{\tt bedaque@mocha.phys.washington.edu}},
  H.-W. Hammer$^{b,}$\thanks{{\tt hammer@triumf.ca}},
  and U. van Kolck$^{c,d,}$\thanks{{\tt vankolck@krl.caltech.edu}} }

\address
{ $^a$~Institute for Nuclear Theory,
 University of Washington, Seattle, WA 98195, USA
~\\$^b$TRIUMF, 4004 Wesbrook Mall,
Vancouver, B.C., Canada V6T 2A3
~\\$^c$ Kellogg Radiation Laboratory, 106-38,
 California Institute of Technology, Pasadena, CA 91125, USA\\
$^d$ Department of Physics,
 University of Washington, Seattle, WA 98195, USA}
\maketitle

\begin{abstract}
We discuss renormalization of the
non-relativistic three-body problem with short-range forces.
The problem becomes non-perturbative at momenta of the
order of the inverse of the
two-body scattering length, and an infinite number of
graphs must be summed.
This summation leads to a cutoff dependence
that does not appear in any order in perturbation theory. We argue
that this cutoff dependence can be absorbed in a single three-body 
counterterm and compute
the running of the three-body force with the cutoff.
We comment on relevance of this result for the
effective field theory program in nuclear and molecular physics.

\end{abstract}

\setcounter{page}{1}
\vskip 0.5cm

Systems composed of particles with momenta $k$ much smaller
than the inverse
range $1/R$ of their interaction are common in nature.
This separation of scales can be exploited
by the method of effective field theory (EFT) that provides a
systematic
expansion in powers of the small parameter $k R$ \cite{gospel}.
Generically, the two-body scattering length $a_2$
is comparable to $R$, and low-density systems with
$k\ll 1/a_2$ can be described to any order in $k R$
by a finite number of EFT graphs \cite{braaten}.
However, there are many interesting systems,
such as those made out of nucleons or of $^4$He atoms,
for which
$a_2$ is much larger than $R$.
In this case the expansion becomes
non-perturbative at momenta of the order of $1/a_2$,
in the sense that an infinite number of diagrams
must be resummed.
This resummation generates a new expansion in powers of $k R$ where
the full dependence in $k a_2$ is kept.
Consequently, the EFT
is valid beyond $k\sim 1/a_2$, comprising,
in particular, bound states of size $\sim a_2$.
There has been
enormous progress recently in dealing with this problem in the two-body case
\cite{edict},
where
the resummation is equivalent to effective range theory \cite{1stooge}.
Ultraviolet (UV) divergences appear in graphs with leading-order interactions
and their resummation contains
arbitrarily high powers of the cutoff.
A crucial point is that
this cutoff dependence
can be absorbed in the coefficients of the leading-order
interactions themselves.  All our ignorance
about the influence of short-distance physics on low-energy phenomena
is then embodied in these few coefficients,
and EFT retains its predictive power.
However, the extension of this program
to three-particle systems presents us with a puzzle \cite{1stoogetoo}.
Although in some fermionic channels the resummed leading two-body
interactions lead to unambiguous
and very successful predictions \cite{2stooges,3stooges},
amplitudes in bosonic systems and other fermionic channels
show sensitivity to the UV cutoff,
as evidenced in the
Thomas \cite{thomas} and Efimov \cite{efimov} effects.
This happens even though each leading-order three-body diagram with
resummed two-body interactions
is individually UV finite.
We will argue below that
the addition of a one-parameter three-body force counterterm at leading
order is necessary and sufficient to eliminate
this cutoff dependence.
This result extends the EFT program to three-particle systems with
large two-body scattering lengths, including the approach of
Ref. \cite{3musketeers} where pions are treated perturbatively.

\begin{figure}[t]
\begin{center}
\epsfxsize=8cm
\centerline{\epsffile{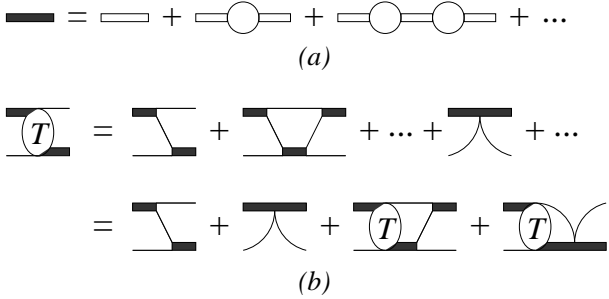}}
\end{center}
\caption{(a) Dressing of the dibaryon. (b) Diagrams contributing to
particle/bound-state scattering.}
\label{fig1}
\end{figure}

The most general Lagrangian involving a non-relativistic
boson $\psi$
and invariant under
small-velocity Lorentz, parity, and time-reversal transformations
is
\begin{equation}
\label{lag}
{\cal L}  =  \psi^\dagger
             (i\partial_{0}+\frac{\vec{\nabla}^{2}}{2M})\psi
 - \frac{C_0}{2} (\psi^\dagger \psi)^2
 - \frac{D_0}{6} (\psi^\dagger\psi)^3 + \ldots ,\nonumber
\end{equation}
where the ellipsis denote terms with more derivatives and/or fields;
those with more fields will not contribute to the
three-body amplitude,
while those with more derivatives are suppressed at low
momentum.
It is convenient \cite{transvestite}
to rewrite this theory
by introducing a dummy field $T$ (called ``dibaryon'' in analogy
to the nuclear case) with quantum
numbers of two bosons,
\begin{eqnarray}
\label{lagt}
{\cal L}  &=&  \psi^\dagger
             (i\partial_{0}+\frac{\vec{\nabla}^{2}}{2M})\psi
         + \Delta T^\dagger T
  -\frac{g}{\sqrt{2}} (T^\dagger \psi\psi +\mbox{h.c.})\nonumber\\
      &+&h T^\dagger T \psi^\dagger\psi
 +\ldots
\end{eqnarray}
\noindent
The arbitrary scale $\Delta$ is included to give
the field $T$ the usual mass dimension.
Observables depend on
the parameters of Eq. (\ref{lagt}) only through the combinations
$C_0\equiv  g^2/\Delta = 4\pi a_2/M$
and $D_0\equiv -3hg^2/\Delta^2$.
The (bare) dibaryon propagator is simply a constant $i/\Delta$,
while the particle propagator reduces to the usual non-relativistic
form $i/(p^0-p^2/2M)$. First, we consider the dressing of the dibaryon in
Fig. \ref{fig1}(a) at a generic momentum $Q$.
The boson loop has a linear UV divergence
that is absorbed in $g^2/\Delta$, a
finite piece
$\sim M g^2 Q/4\pi \Delta^2$  determined by the unitarity cut,
and terms suppressed by powers
of $Q$ over the cutoff $\Lambda$
that are subleading
and of the same size as terms in Eq. (\ref{lagt}) already disregarded.
Using the relation between $g$ and $a_2$,
we see that the finite piece is
$\sim Qa_2/\Delta$ and, consequently, has to be resummed
to all orders for $Q\simge 1/a_2$.
As a result, the dressed dibaryon propagator to leading order is given by
\begin{equation}
\label{Dprop}
i S(p) =  \frac {1}{- \Delta
             + \frac{M g^{2}}{4\pi}
               \sqrt{-M p^0+\frac{\vec{p}^{\,2}}{4}-i\epsilon} +i\epsilon} .
\end{equation}
\noindent
Attaching
four boson lines to this propagator gives the
two-particle scattering amplitude at leading order, which is identical
to the effective range expansion at the order of the scattering
length. Further corrections give the next terms in the effective range
expansion \cite{1stooge}.

Let us now
consider
particle/bound-state scattering.
All diagrams contributing to this process in leading order
are illustrated in Fig. \ref{fig1}(b).
Each of the diagrams including only two-body interactions
gives a contribution
of order $\sim M g^2/Q^2\sim 4\pi \Delta a_2^3$. (The
properly normalized amplitude is independent of
the arbitrary parameter $\Delta$; it appears here only because of
our choice of interpolating field for the bound state.)
The relative size of graphs that include a three-body force
will be discussed shortly.

The sum of all the diagrams in Fig. \ref{fig1}(b) can be accomplished
by solving the equation
represented by the second equality in Fig. \ref{fig1}(b)
\cite{1stoogetoo,2stooges,3stooges,skorny}:

\begin{equation}
\label{aeq}
a(p)
=K(p,k)+\frac{2\lambda}{\pi}\int_0^\Lambda dq\ K(p,q)
\frac{q^2}{q^2-k^2-i\epsilon} a(q),
\end{equation}
\noindent
where $k$ ($p$) is the incoming (outgoing) momentum,
$M E = 3k^2/4 - 1/a_2^2$
is the total energy, $a(p=k)$ is the scattering amplitude normalized
in such a way that $a_3=-a(0)$ is the particle/bound-state scattering
length, and
\begin{eqnarray}
K(p,q)= \frac{4}{3}& &(\frac{1}{a_2}+\sqrt{\frac{3}{4}p^2-M E})\\
   & &\left[\frac{1}{pq}{\rm ln}
    \left(\frac{q^2+p q+p^2-M E}
               {q^2-q p+p^2-M E}\right)
    +\frac{h}{Mg^2} \right]\nonumber.
\end{eqnarray}
\noindent
The parametric dependence of $a(p)$ on $k$ is
kept implicit. We are interested in $\lambda=1$ for the boson case.
This equation reduces to the expressions found in
Ref. \cite{1stoogetoo,skorny}
when $h=0$.
Three nucleons in the spin $J=1/2$ channel obey a pair
of integral equations with similar properties to this bosonic equation,
while the spin $J=3/2$ channel corresponds to $\lambda=-1/2$.

\begin{figure}[tbh]
\begin{center}
\epsfxsize=8cm
\centerline{\epsffile{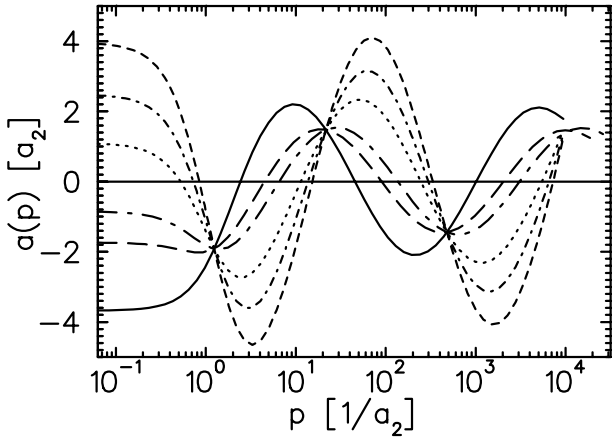}}
\end{center}
\caption{$a(p)$ for $k=0$. Full, dashed, and dash-dotted curves are
for $H=0$ and $\Lambda=1.0,\,2.0,\,3.0\times 10^4 a_2^{-1}$, respectively.
Dotted, short-dash-dotted, and short-dashed curves are for
$\Lambda=10^4 a_2^{-1}$ and $H=-6.0,\,-2.5,\,-1.8$, respectively.}
\label{fig2}
\end{figure}

Let us look at the asymptotic behavior of the solution of
Eq. (\ref{aeq}) in the case $h=0$.
For $1/a_2\ll p\ll \Lambda $ (but $k\sim 1/a_2$), the integral in
Eq. (\ref{aeq}) is dominated by momenta in the intermediate region
$1/a_2\ll q\ll \Lambda $ and the equation
\begin{equation}
\label{asphieq}
a(p)= \frac{4 \lambda}{\sqrt{3} \pi}
                   \int_0^\infty \: \frac{dq}{q}\
                   a(q)\, {\rm ln} \left(\frac{q^2+pq+p^2}{q^2-pq+p^2}\right),
\end{equation}
\noindent
holds
up to terms suppressed by powers of
$1/p a_2$ and/or $p/\Lambda$.
The scale invariance of Eq. (\ref{asphieq}) suggests an
ansatz of the form $a(p)\sim p^s$, which works if $s$ satisfies
\begin{equation}
\label{seq}
1- \frac{8\lambda}{\sqrt{3}}
   \frac{\sin\frac{\pi s}{6}}{s \cos\frac{\pi s}{2}}=0.
\end{equation}
\noindent
The solutions of Eq. (\ref{seq}) come in pairs due to the additional
symmetry $a(p)\rightarrow a(1/p)$ of Eq. (\ref{asphieq}).
For $\lambda<\lambda_c={3 \sqrt{3}\over 4 \pi}\simeq 0.4135$, Eq.
(\ref{seq}) has only real roots.
However,
for $\lambda =1$ there are two
imaginary solutions $s=\pm i s_0$, with $s_0\simeq 1.0064$.
Both make the
integral in Eq. (\ref{aeq}) UV finite and are equally acceptable:
$a(p)$ is given in the intermediate region by a linear combination
of $p^{i s_0}$ and $p^{-i s_0}$.
Eq. (\ref{asphieq}) is homogeneous so it clearly cannot
determine the overall normalization of $a$,
but we now see that it cannot uniquely determine
the phase either. However, Eq. (\ref{asphieq}) with finite $\Lambda$
has a solution with a well determined phase
that in the intermediate region $1/a_2\ll p\ll \Lambda$
is, on dimensional grounds,

\begin{equation}
a(p)=A \cos(s_0 \ln \frac{p}{\Lambda} + \delta), \label{asym}
\end{equation}
\noindent
where $\delta$ is some dimensionless, cutoff-independent number.
The limit
$\Lambda\rightarrow\infty$ is not well defined
because Eq. (\ref{aeq}) does not have a unique solution
in this limit. (A rigorous proof that Eq. (\ref{aeq}) does not have a
unique solution can be found in
\cite{danilov}.) This non-uniqueness comes from our idealization 
of the interactions as point-like. 
Note that subleading contributions from the integration range
$\Lambda \leq q\leq \infty$ change phase and amplitude significantly.
Numerical solutions of
Eq. (\ref{aeq}) with $k=0$ for different values of $\Lambda$
are plotted in Fig. \ref{fig2}.
(Our results agree with those in Ref. \cite{kharchenko}
for the appropriate cutoff values.)
We observe that indeed the behavior
of $a(p)$ in the region $1/a_2\ll p \ll \Lambda$ is given by Eq.
(\ref{asym}) and that small differences in
the asymptotic phase lead to large differences in the particle/bound-state
scattering length.

Note that if the series of diagrams in Fig. \ref{fig1}(b) was truncated 
at some finite number of loops one would miss 
the correct asymptotic behavior of $a(p)$ (Eq. (\ref{asym})) that
generates the cutoff dependence.
This is because $s_0$ (and its expansion in powers of $\lambda$)
vanish in a neighborhood of $\lambda=0$
and the truncation of the series in  Fig. \ref{fig1}(b)
is equivalent to perturbation theory in $\lambda$.

This cutoff dependence comes from the behavior of the amplitude in
the UV region, where the EFT Lagrangian, Eq. (\ref{lagt}), is not
to be trusted.
When the low-energy expansion is perturbative, the cutoff-dependent
contribution  from high loop momenta can be expanded in powers of
the low external momenta and cancelled by local terms in the Lagrangian.
Thus all uncertainty coming
from the high momentum behavior of the theory is
parametrized by a few local counterterms.
The present case is complicated by the non-analytic cutoff
dependence of the amplitude around $p=0$.
That, however, does not mean that the renormalization program
is doomed: a three-body force term of sufficient strength
contributes not only at tree level, but also in loops
dressed by any number of two-particle interactions. This 
generates non-local contributions precisely of the same
form as the cutoff-dependent terms generated by the two-body force alone.
To see how that comes about we turn on the three-body force term  and
write $h(\Lambda)=2M g^2 H(\Lambda)/\Lambda^2$ assuming $H(\Lambda)\sim 1$.
The asymptotic Eq. (\ref{asphieq}) becomes
\begin{equation}
\label{aseqH}
a(p)= \frac{4}{\sqrt{3} \pi}
                   \int_0^\Lambda \: \frac{dq}{q}\ a(q)
                \left[ {\rm ln} \left(\frac{q^2+ pq+p^2}{q^2-p q+p^2} \right)
                    + 2 H\frac{p q}{\Lambda^2}\right],
\end{equation}
\noindent
where we have set $\lambda=1$ for definiteness.
For $p\sim \Lambda$ the
term proportional to $H$ becomes important and
$a(p\sim \Lambda)$ has a complicated form.
In the range $1/a_2\ll p\ll \Lambda$, however, the three-body force
is suppressed by $p/\Lambda$ compared to the logarithm
and can be disregarded. Consequently,
Eq. (\ref{asym}) is still correct in the intermediate region.
The effect of a
finite value of $H$ can be at most to change the values of the amplitude
$A$ and the phase $\delta$, which become
functions of $H$.
As shown in Fig. \ref{fig2}, this is confirmed by
numerical solutions:
while different values of the three-body force preserve the form of the
solution, the phase (and amplitude)  are changed.
If $H$ is chosen to be a function of $\Lambda$
such as to cancel the explicit $\Lambda$ dependence,
we can make the solution of Eq. (\ref{aeq})
cutoff independent for all $p\ll \Lambda$.
In particular the
scattering amplitude that is determined by the on-shell value
$a(k)$ with $k\sim 1/a_2$ will be cutoff independent as well.
For this to be possible $A$ and $\delta$ must depend on
the same combination of $\Lambda$ and $H$.
Thus $H(\Lambda)$ must be chosen such that
\begin{equation}
\label{Lbar}
-s_0 \ln \Lambda + \delta(H(\Lambda))= -s_0 \ln \Lambda_\star ,
\end{equation}
\noindent
where $\Lambda_\star$ is a parameter fixed by experiment or by matching
with a microscopic model.

We can get a handle on the form of $H(\Lambda)$ by
considering Eq. (\ref{aeq}) with two different values of the cutoff
$\Lambda$ and $\Lambda '>\Lambda$, whose solutions we denote by $a(p)$ and
$a'(p)$.
In the intermediate region $1/a_2\ll p\ll \Lambda$
the equations for $a(p)$ and $a'(p)$ will have the same form except for
\begin{equation}
\frac{2}{\pi}\! \left[ \int_\Lambda^{\Lambda'} \!\!\!\! dq \, K'(p,q) \, a'(q) 
+2 \! \! \int_\mu^{\Lambda} \! \! \! dq
\left( \frac{H(\Lambda')}{\Lambda'}-\frac{H(\Lambda)}{\Lambda} \right)a'(q)
\right]
\label{diff}
\end{equation}
\noindent
where $\mu \ll \Lambda$ is an arbitrary scale,
and we dropped
terms suppressed  by  further powers of $1/pa_2$ and $p/\Lambda$.
Assuming $a'(p)$ has the same phase $\cos(s_0 \ln(p/\Lambda_\star))$
as $a(p)$ even
for $p\sim\Lambda'$, we can make the terms in Eq. (\ref{diff})
vanish by choosing
\begin{equation}
\label{h}
H(\Lambda)= -
                   \frac{\sin(s_0\ln({\Lambda}/{\Lambda_\star})-
                   {\rm arctg}(1/s_0))}
                 {\sin(s_0 \ln({\Lambda}/{\Lambda_\star})+
                   {\rm arctg}(1/s_0))}.
\end{equation}
\noindent
Since $K'-K$ nearly vanishes for all $p\ll \Lambda$, $a'(p)$ has also the same
amplitude as $a(p)$ in the intermediate region.
That is, with $H(\Lambda)$ chosen like Eq. (\ref{h}), $a(p)=a'(p)$
for all values $p\ll\Lambda$ (up to terms suppressed
by $p/\Lambda$), and the on-shell amplitude
$a(k)$ for $k\ll\Lambda$ will be $\Lambda$ independent.
Once the parameter $\Lambda_\star$
is fitted to an experimental datum at
a certain energy, the energy dependence can be predicted.

\begin{figure}[t]
\begin{center}
\epsfxsize=8cm
\centerline{\epsffile{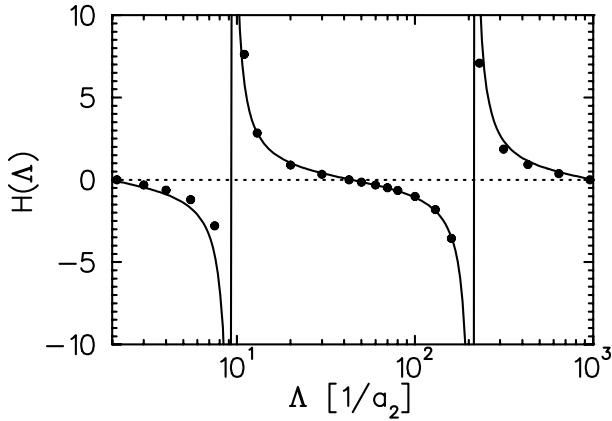}}
\end{center}
\caption{Three-body force as a function of the cutoff $\Lambda$:
numerical solution (dots) and Eq.(\ref{h}).}
\label{fig3}
\end{figure}

\begin{figure}[t]
\begin{center}
\epsfxsize=8cm
\centerline{\epsffile{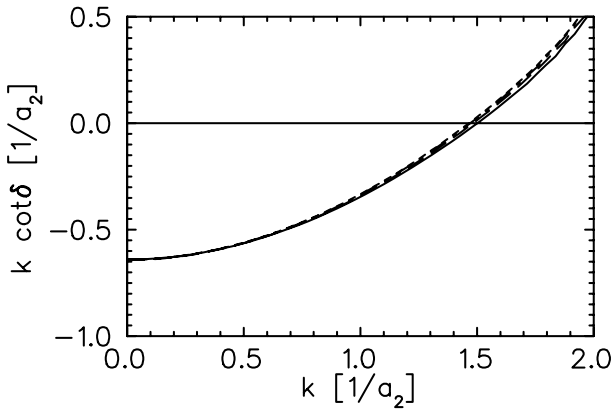}}
\end{center}
\caption{Energy dependence: $k\cot\delta$ for different
cutoffs ($\Lambda=42.6,\,100.0,\,230.0,\,959.0\times a_2^{-1}$).}
\label{fig4}
\end{figure}

We also determine $H(\Lambda)$ numerically by finding the value
of $H$ that keeps the scattering length $a_3=-a(0)$
constant for each value of $\Lambda$ varying over a large range.
These values are plotted as a function of $\Lambda$
in Fig. \ref{fig3} together with
$H(\Lambda)$
given by Eq. (\ref{h}).
For illustration we used $a_3= 1.56 a_2$, but have verified that similar
agreement holds for other values of $a_3$.
In Fig. \ref{fig4} we show the corresponding $k \cot\delta=i k +a(k)^{-1}$,
where $\delta$ is the $S$-wave phase shift for particle/bound-state
scattering, for several values of $\Lambda$.
As argued above, it is insensitive to
$\Lambda$ as long as $k\ll \Lambda$.
The effective range, for example, is predicted as $r_3= 0.57 a_2$.
Note that the three-body force
discussed here is not the one used in realistic potential models where 
the effective cutoff is at much higher scales.

These arguments hold for the bound-state problem as well
because the inhomogeneous terms played no role.
We have solved the homogeneous equation with
the $H(\Lambda)$ of Fig. \ref{fig3}.
Only the shallowest bound state is large enough to
be unequivocally within the limits of applicability of the EFT;
it has a cutoff-independent
binding energy of $B_3=1.5/M a_2^2$.

The value for the ratio $a_3/a_2$ used above
is the one suggested
by the values of
$a_2=124.7$ \AA \ and $a_3=195$ \AA \
given by a phenomenological $^4$He-$^4$He potential \cite{heliumpot} consistent
with the recent measurement of the dimer binding energy  \cite{dimer}.
Fig. \ref{fig4} then represents the
phase shifts for atom/dimer scattering, with an effective
range $r_3= 71$ \AA.
Similarly, our result for the shallowest bound state
suggests an excited state of the trimer at $B_3=1.2$ mK.
Because the integral equations are similar,
our arguments are relevant for three-fermion systems
with internal quantum numbers as well \cite{more3stooges}.
The approach of Ref. \cite{3musketeers} then suggests
that our bound-state results would provide a reasonable
estimate of the triton binding energy.

In conclusion,
we have provided analytical and numerical evidence that
renormalization of the three-body problem with short-range forces
requires in general
the presence of a one-parameter contact three-body force in leading order.
This opens up the possibility of applying the EFT method to a large class
of systems of three or more particles with short-range forces.


We thank V. Efimov, H. M{\"u}ller, and D. Kaplan for 
helpful discussions.
HWH acknowledges 
hospitality of the Nuclear Theory Group and the INT in Seattle.   
Research supported in part by the U.S. DOE
(DOE-ER-40561 and DE-FG03-97ER41014),
the NSERC of Canada,
and the U.S. NSF 
(PHY94-20470).


\end{document}